\documentclass[12pt,prb,aps]{revtex4-1}
\usepackage{fullpage}
\usepackage{amssymb,amsmath}
\allowdisplaybreaks[1]
\usepackage{epsf}

\newcommand {\ltapp} {\stackrel {_{\normalsize<}}{_{\normalsize \sim}}}

\begin{document}
\title{Effect of Nonlinear Energy Transport on Neoclassical Tearing Mode Stability in Tokamak Plasmas\\[0.5ex]
~\\[0.5ex]
{\rm Richard Fitzpatrick}\\[0.5ex]
{\it Institute for Fusion Studies}\\
{\it Department of Physics}\\
{\it University of Texas at Austin}\\{\it Austin, TX 78712}\\~}

\begin{abstract}
An investigation is made into the effect of the reduction in anomalous perpendicular electron heat transport inside
the separatrix of a magnetic island chain associated with a neoclassical tearing mode in a tokamak plasma, due to the flattening of the electron temperature
profile in this region, on the overall stability of the mode. The onset of the neoclassical
tearing mode is governed by the ratio of the divergences of the parallel and perpendicular electron heat fluxes in the vicinity of the island chain. By increasing the degree of transport reduction, the onset of the mode, as the divergence ratio is gradually increased, can be made more and more
abrupt. Eventually, when the degree of transport reduction passes a certain critical value, the onset of the neoclassical tearing mode becomes
discontinuous. In other words, when some critical value of the divergence ratio is reached, there is a sudden bifurcation to a branch of neoclassical tearing mode solutions. Moreover, once this bifurcation has been triggered, the divergence ratio must reduced by a substantial factor to
trigger the inverse bifurcation. 
\end{abstract}
\maketitle

\section{Introduction}
Neoclassical tearing modes are large-scale magnetohydrodynamical instabilities that cause the axisymmetric, 
toroidally-nested, magnetic flux-surfaces of a tokamak plasma to reconnect to form helical magnetic island structures on low mode-number rational
magnetic flux surfaces.\cite{wesson} Island formation leads to a degradation of  plasma energy confinement.\cite{chang} Indeed, the confinement degradation associated with neoclassical tearing modes constitutes  a major
impediment to the development of effective operating scenarios in present-day and future tokamak experiments.\cite{sauter}
Neoclassical tearing modes are driven by the flattening of the temperature and density   profiles within
the magnetic separatrix of the associated island chain,  leading to the suppression of the neoclassical
bootstrap current in this region, which has a destabilizing effect on the mode.\cite{car} The degree of
 flattening of a given profile (i.e., either the density, electron temperature, or ion temperature profile) within the island separatrix depends on the ratio of the associated perpendicular (to the magnetic field)
and parallel transport coefficients.\cite{rf}

The dominant contribution to the perpendicular
transport in tokamak plasmas comes from small-scale drift-wave turbulence, driven by plasma density and temperature gradients.\cite{wesson}  
The fact that the density and temperature profiles are flattened within the magnetic
separatrix of a   magnetic island chain implies a substantial reduction in the  associated perpendicular transport coefficients 
in this region. Such a reduction has been observed in gyrokinetic simulations,\cite{sim1,sim2,sim3,sim4} as well as in experiments.\cite{ex1,ex2,ex3,ex4,ex5}
A strong reduction in perpendicular transport within the magnetic separatrix calls into question the conventional analytic theory of neoclassical tearing modes in which the perpendicular transport coefficients are assumed to 
spatially uniform in the island region.\cite{rf} 

The aim of this paper is to investigate the effect of the reduction in perpendicular transport inside the
separatrix  of a neoclassical magnetic island chain, due to profile flattening in this region, on the overall stability of the mode. For the sake of simplicity, we
shall only consider the influence of the flattening of the electron temperature profile on mode stability. However, the analysis
contained in this paper could be generalized, in a fairly straightforward manner, to take into account the influence of the
flattening of the ion temperature and density profiles. 

\section{Preliminary Analysis}\label{s2}
\subsection{Fundamental Definitions}
Consider a large aspect-ratio, low-$\beta$, circular cross-section,
 tokamak plasma equilibrium. Let us adopt a right-handed cylindrical coordinate system ($r$, $\theta$, $z$) whose symmetry axis ($r=0$) coincides with the
magnetic axis of the plasma. The system is assumed to be periodic in the $z$-direction with period $2\pi\,R_0$, where $R_0$ is the simulated major
plasma radius. It is helpful to define the simulated toroidal angle $\varphi=z/R_0$. 
  The coordinate $r$  serves as a label
for the unperturbed (by the tearing mode) magnetic flux-surfaces. Let the equilibrium toroidal magnetic field, $B_z$, and the equilibrium toroidal plasma current both run in the $+z$ direction. 

 Suppose that a neoclassical tearing mode generates a helical magnetic island chain,
with $m_\theta$ poloidal periods, and $n_\varphi$ toroidal periods, that is embedded in
 the aforementioned plasma. The island chain is assumed to be  radially localized in the vicinity of its
associated 
rational surface, minor radius $r_s$,  which is defined as the unperturbed magnetic flux-surface at which $q(r_s)=m_\theta/n_\varphi$. Here, $q(r)$ is the   
safety-factor profile (which is assumed to be a monotonically increasing function of $r$). 
Let the full radial width of the island chain's magnetic separatrix be $W$.
In the following, it is assumed that
$\epsilon_s \equiv r_s/R_0\ll 1$ and $W/r_s\ll 1$. 

It is convenient to employ a frame of reference that co-rotates with the magnetic island chain. All fields  are assumed to depend (spatially) only on the radial
coordinate $r$ and the helical angle $\zeta=m_\theta\,\theta-n_\varphi\,\varphi$. Let $k_\theta=m_\theta/r_s$, $q_s=m_\theta/n_\varphi$, and $s_s=d\ln q/d\ln r|_{r_s}$. The magnetic shear length at the rational
surface is defined $L_s= R_0\,q_s/s_s$. Moreover, the unperturbed (by the magnetic island) electron temperature
gradient scale-length at the rational surface takes the form $L_T = -1/(d\ln T_0/dr)_{r_s}$, where $T_0(r)$
is the unperturbed electron temperature profile.  In the following, it is assumed that $L_T>0$, as is generally
the case in conventional tokamak plasmas.\cite{wesson} 

The helical magnetic flux is defined
\begin{equation}
\chi(x,\zeta) = -\frac{B_z}{R_0}\int_0^x \left(\frac{1}{q}-\frac{1}{q_s}\right) (r_s+x)\,dx + \delta \chi(x,\zeta),
\end{equation}
where $x=r-r_s$, and  the magnetic field perturbation associated with the tearing mode is written $\delta {\bf B}=\nabla \times (\delta\chi\,{\bf e}_z)$. It is easily demonstrated that ${\bf B}\cdot\nabla \chi=0$,
where ${\bf B}$ is the total magnetic field.\cite{ruth} Hence, $\chi$ is a magnetic flux-surface label. 
It is helpful to introduce the normalized helical magnetic flux, $\psi = (L_s/B_z\,w^{\,2})\,\chi$, where
$w=W/4$. The normalized flux in the vicinity of the rational surface is assumed to
take the form\,\cite{ruth}
\begin{equation}
\psi(X,\zeta) = \frac{1}{2}\,X^{\,2} + \cos\zeta,
\end{equation}
where $X=x/w$. 
As is well-known, the contours of $\psi$ map out a symmetric (with respect to $X=0$), constant-$\psi$,\cite{fkr} magnetic island chain whose O-points lie at $\zeta=\pi$, $X=0$,
and $\psi=-1$, and whose X-points lie at $\zeta=0$, $2\pi$, $X=0$, and $\psi=+1$. The chain's magnetic separatrix
corresponds to $\psi=+1$, the region inside the separatrix to $-1\leq \psi<1$, and the region outside the separatrix to $\psi>1$. The
full radial width of the separatrix (in $X$) is 4. 

Finally, the electron  temperature profile in the vicinity of the rational surface is written
\begin{equation}
T(X,\zeta)=T_s\left[1-\left(\frac{w}{L_T}\right)\delta T(X,\zeta)\right],
\end{equation}
where $T_s=T_0(r_s)$, and 
\begin{equation}\label{e4}
\left.\delta T(X,\zeta)\right|_{\lim |X|\rightarrow\infty} = X.
\end{equation}
Note that $\delta T(X,\zeta)$ is an odd function of $X$. In the following, it is assumed that $w/L_T\ll 1$. 

\subsection{Electron Energy Conservation Equation}
The steady-state electron temperature profile in the vicinity of the island chain is governed by the following well-known electron energy conservation equation:\,\cite{brag,rf}
\begin{equation}\label{e1}
\left(\frac{W}{W_c}\right)^{4}\left[\left[\delta T,\psi\right],\psi\right] + \frac{\partial^{\,2} \delta T}{\partial X^{\,2}} =0,
\end{equation}
where
\begin{equation}
[A,B] \equiv \frac{\partial A}{\partial X}\,\frac{\partial B}{\partial \zeta}-\frac{\partial A}{\partial\zeta}\,\frac{\partial B}{\partial X},
\end{equation}
and
\begin{equation}
W_c =4\left(\frac{\kappa_\perp}{\kappa_\parallel}\right)^{1/4}\left(\frac{L_s}{k_\theta}\right)^{1/2}.
\end{equation}
Here, $\kappa_\perp$ and $\kappa_\parallel$ are the perpendicular (to the magnetic field) and
parallel electron thermal conductivities, respectively. 
The first term on the right-hand side of Eq.~(\ref{e1}) represents the divergence of the parallel (to the magnetic field)
electron heat flux, whereas the second term represents the divergence of the perpendicular electron heat flux. [In fact, because 
$[[\delta T,\psi],\psi]$ and $\partial^{\,2} \delta T/\partial X^{\,2}$ are both ${\cal O}(1)$ in our normalization scheme, the ratio of the divergences of the parallel and perpendicular heat fluxes is effectively measured by $(W/W_c)^4$.]
The
quantity $W_c$ is the critical island width above which the former term dominates the latter, causing
the temperature profile to flatten within the island separatrix.\cite{rf} In writing Eq.~(\ref{e1}), we have neglected any
localized sources or sinks of heat in the island region. We have also assumed that $\kappa_\perp$ and
$\kappa_\parallel$ are spatially uniform in the vicinity of the rational surface. The latter assumption is relaxed
in Sect.~\ref{s3}

\subsection{Narrow-Island Limit}\label{snarrow}
Consider the so-called {\em narrow-island limit}\/ in which $W\ll W_c$.\cite{rf} Let
\begin{equation}
Y = \left(\frac{w}{w_c}\right)X.
\end{equation}
Equation~(\ref{e1}) transforms to give
\begin{align}
\frac{\partial ^{\,2} \delta T}{\partial Y^{\,2}}+Y^{\,2}\,\frac{\partial^{\,2} \delta T}{\partial \zeta^{\,2}}&=
-\left(\frac{W}{W_c}\right)^{\,2}
\left(\sin\zeta\,\frac{\partial \,\delta T}{\partial\zeta} + 2\,Y\,\sin\zeta\,\frac{\partial^{\,2} \delta T}{\partial Y\,\partial\zeta}+Y\,\cos\zeta\,\frac{\partial\, \delta T}{\partial Y}\right)
\nonumber\\[0.5ex]&\phantom{=}-\left(\frac{W}{W_c}\right)^{\,4} \sin^2\zeta\,\frac{\partial^{\,2} \delta T}{\partial Y^{\,2}}.
\end{align}
We can write 
\begin{equation}\label{e11}
\delta T(Y,\zeta)  = \left(\frac{W_c}{W}\right)Y + \left(\frac{W}{W_c}\right) T_1(Y,\zeta) +
 {\cal O}\left(\frac{W}{W_c}\right)^{3},
\end{equation}
where 
\begin{equation}
\frac{\partial ^{\,2} T_1}{\partial Y^{\,2}}+Y^{\,2}\,\frac{\partial^{\,2} T_1}{\partial \zeta^{\,2}}=-Y\,\cos\zeta,
\end{equation}
subject to the boundary conditions $T_1(0,\zeta)=0$, and $T_1\rightarrow 0$ as $|Y|\rightarrow\infty$. 
Note that the solution (\ref{e11}) automatically satisfies the boundary condition (\ref{e4}). 
It follows that 
\begin{equation}
T_1 (Y,\zeta) = \frac{\sqrt{2}}{4}\,f\left(\sqrt{2}\,Y\right)\,\cos\zeta,
\end{equation}
where $f(p)$ is the well-behaved solution of 
\begin{equation}\label{e13}
\frac{d^{\,2}f}{dp^{\,2}}-\frac{1}{4}\,p^{\,2}\,f = -p
\end{equation}
that satisfies $f(0)=0$, and $f\rightarrow 0$ as $|p|\rightarrow\infty$.  Note that $f(-p)=-f(p)$. 
Hence, in the narrow-island limit,\cite{rf} 
\begin{equation}\label{e14}
\delta T(X,\zeta) = X + \left(\frac{W}{W_c}\right)\frac{\sqrt{2}}{4}\,f\left(\sqrt{2}\,\frac{W}{W_c} X\right)\,\cos\zeta  +{\cal O}\left(\frac{W}{W_c}\right)^{3}.
\end{equation}

\subsection{Wide-Island Limit}\label{swide}
Consider the so-called {\em wide-island limit}\/ in which $W\gg W_c$.\cite{rf} We can write
\begin{equation}\label{e16}
\delta T (X,\zeta)=\bar{T}(\psi) + \left(\frac{W_c}{W}\right)^{\,4}\tilde{T}(\psi,\zeta),
\end{equation}
where $\bar{T}$ and $\tilde{T}$ are both ${\cal O}(1)$, and 
\begin{equation}
\langle\tilde{T}\rangle = 0.
\end{equation}
Here, $\langle\cdots\rangle$ is the so-called {\em flux-surface average operator}.\cite{ruth} This operator is defined as follows:
\begin{equation}\label{e68}
\langle A\rangle = \int_{\zeta_0}^{2\pi-\zeta_0}\frac{A_+(\psi,\zeta)}{\sqrt{2\,(\psi-\cos\zeta)}}\,\frac{d\zeta}{2\pi}
\end{equation}
for $-1\leq \psi\leq  1$, and 
\begin{equation}\label{e69}
\langle A\rangle = \int_0^{2\pi}\,\frac{A(s,\psi,\zeta)}{\sqrt{2\,(\psi-\cos\zeta)}}\,\frac{d\zeta}{2\pi}
\end{equation}
for $\psi>1$, where $s={\rm sgn}(X)$, $\zeta_0=\cos^{-1}({\mit\psi})$, and
\begin{equation}\label{e70}
A_+(\psi,\zeta) = \frac{1}{2}\left[A(+1,\psi,\zeta) +A(-1,\psi,\zeta)\right].
\end{equation}
Note that $\langle [A,\psi]\rangle \equiv 0$ for all $A$. 

Equations~(\ref{e1}) and (\ref{e16}) can be combined to give 
\begin{equation}\label{e79}
[[\tilde{T},\psi],\psi] + \left(\frac{W_c}{W}\right)^4 \frac{\partial^{\,2} \tilde{T}}{\partial X^{\,2}} + \frac{\partial^{\,2}\bar{T}}{\partial X^{\,2}}=0.
\end{equation}
The flux-surface average of the previous equation yields 
\begin{equation}
\left\langle\frac{\partial^{\,2}\bar{T}}{\partial X^{\,2}}\right\rangle ={\cal O}\left(\frac{W_c}{W}\right)^4,
\end{equation}
which implies that 
\begin{equation}
\frac{d}{d\psi}\left(\langle X^{\,2}\rangle\,\frac{d\bar{T}}{d\psi}\right) \simeq 0.
\end{equation}
The previous equation can be integrated to give
\begin{equation}\label{e23}
\bar{T}(\psi) 
=\left\{
\begin{array}{ccc}
0&\mbox{\hspace{1cm}} & -1\leq \psi \leq 1\\[0.5ex]
s\int_1^\psi \frac{d\psi'}{\langle X^{\,2}\rangle (\psi')}&&\psi >1
\end{array}
\right.,
\end{equation}
which satisfies the boundary condition (\ref{e4}). 
Hence,  in the wide-island limit,\cite{rf}
\begin{equation}\label{e24}
\delta T(X,\zeta) =\bar{T}(\psi) +  {\cal O}\left(\frac{W_c}{W}\right)^{\,4}.
\end{equation}

\subsection{Modified Rutherford Equation}
The temporal evolution of the island width is governed by the so-called {\em modified Rutherford  equation}, which  takes the form\,\cite{ruth,rf,car}
\begin{equation}\label{e26}
G_1\,\tau_R\,\frac{d}{dt}\!\left(\frac{W}{r_s}\right) = {\mit\Delta}'\,r_s + G_2\,\alpha_b\,\frac{L_s}{L_T}\,\frac{r_s}{W},
\end{equation}
where
\begin{align}\label{e26a}
G_1 &= 2\int_{-1}^\infty \frac{\langle \cos\zeta\rangle^{\,2}}{\langle 1\rangle}\,d\psi,\\[0.5ex]
G_2 &= 16\int_{-1}^\infty \left\langle \frac{\partial T}{\partial X}\right\rangle \frac{\langle \cos\zeta\rangle}{\langle 1\rangle}\,d\psi.\label{e27}
\end{align}
Here, $\tau_R=\mu_0\,r_s^{\,2}/\eta(r_s)$ is the resistive evolution timescale at the rational surface, and $\eta(r)$ is the
unperturbed plasma resistivity profile. Moreover, ${\mit\Delta}'<0$ is the standard linear tearing stability index.\cite{fkr} 
Finally, $\alpha_b = f_s\,(q_s/\epsilon_s)\,\beta$, where $f_s=1.46\,\epsilon_s^{\,1/2}$ is the fraction of trapped electrons, $\beta=
\mu_0\,n_s\,T_s/B_z^{\,2}$, and $n_s$ is the unperturbed electron number density at the
rational surface. The second term on the right-hand side of Eq.~(\ref{e26}) parameterizes the destabilizing 
influence of the perturbed bootstrap current.\cite{rf,car} Note that, in this paper, for the sake of simplicity, 
we have employed the so-called {\em lowest-order asymptotic matching scheme}\/ described in Ref.~\onlinecite{rf1}.
This accounts for the absence of higher-order island saturation terms in Eq.~(\ref{e26}). 

\section{Effect of Temperature Flattening}\label{s3}
\subsection{Introduction}
In conventional tokamak plasmas, the dominant contribution to the perpendicular electron thermal conductivity, $\kappa_\perp$, 
comes from small-scale  drift-wave turbulence driven by electron temperature gradients.\cite{wesson} The fact that the electron temperature gradient is flattened within the magnetic
separatrix of a sufficiently wide  magnetic island chain implies a substantial reduction in the  perpendicular electron thermal conductivity
in this region. There is clear experimental evidence that this is indeed the
case.\cite{ex2,ex3,ex5} In particular, Ref.~\onlinecite{ex5} reports a reduction in $\kappa_\perp$ at the
O-point of the magnetic island chain associated with a typical neoclassical tearing mode by 1 to 2 orders of magnitude.
Obviously, such a strong reduction in $\kappa_\perp$ within the magnetic separatrix calls into question the conventional analytic model of neoclassical tearing modes, described in Sect.~\ref{s2}, in which $\kappa_\perp$ is assumed to 
spatially uniform in the vicinity of the rational surface. 

\subsection{Nonuniform Perpendicular Electron Conductivity Model}\label{sxx}
As a first attempt to model the reduction in $\kappa_\perp$ due to temperature flattening within
the magnetic separatrix of a neoclassical island chain, let us
write
\begin{equation}
\kappa_\perp = \left\{
\begin{array}{ccc} 
\kappa_{\perp\,1}&\mbox{\hspace{1cm}}& -1\leq \psi\leq 1\\[0.5ex]
\kappa_{\perp\,0} &&\psi > 1
\end{array}
\right.,
\end{equation}
where $\kappa_{\perp\,1}$ and $\kappa_{\perp\,0}$ are spatial constants, with $\kappa_{\perp\,1}\leq\kappa_{\perp\,0}$. Since the mean temperature gradient outside the separatrix of a  neoclassical magnetic island
chain is  similar in magnitude to the equilibrium temperature gradient [see Eq.~(\ref{e42})], it is reasonable to assume that $\kappa_{\perp\,0}$ is
equal to the local (to the rational surface) perpendicular electron thermal conductivity in the absence of an island chain. 

Let
\begin{align}
W_{c\,0} &=4\left(\frac{\kappa_{\perp\,0}}{\kappa_\parallel}\right)^{1/4}\left(\frac{L_s}{k_\theta}\right)^{1/2},\label{e29}\\[0.5ex]
W_{c\,1} &=4\left(\frac{\kappa_{\perp\,1}}{\kappa_\parallel}\right)^{1/4}\left(\frac{L_s}{k_\theta}\right)^{1/2},
\end{align}
be the critical island widths outside and inside the separatrix, respectively. Likewise, let
\begin{align}
\xi_0 &= \left(\frac{W}{W_{c\,0}}\right)^4,\label{e31}\\[0.5ex]
\xi_1 &= \left(\frac{W}{W_{c\,1}}\right)^4,\label{e32}
\end{align}
measure the ratios of the divergences of the parallel and perpendicular electron heat fluxes outside and inside the separatrix,
respectively. Finally, let the parameter 
\begin{equation}\label{e33}
\lambda = \frac{\kappa_{\perp\,1}}{\kappa_{\perp 0}} =\frac{\xi_0}{\xi_1}
\end{equation}
measure the relative reduction of perpendicular electron heat transport within  the island separatrix. 

Let us adopt the following simple model:
\begin{equation}\label{e34}
\lambda = {\rm e}^{-\xi_1}+ \left(1-{\rm e}^{-\xi_1}\right)\delta,
\end{equation}
where $0<\delta\leq 1$. According to this model, the degree of perpendicular transport reduction within the
separatrix is controlled by the parameter $\xi_1$, which measures ratio of the divergences of the
parallel and perpendicular electron heat fluxes {\em inside}\/ the separatrix. (See Sects.~\ref{snarrow} and \ref{swide}.)  If $\xi_1$ is much less than unity then there is no temperature flattening
within the separatrix, which implies that $\lambda =1$ (i.e., there is no reduction in transport). On the other hand,
if $\xi_1$ is much greater than unity then the temperature profile is completely flattened inside the
separatrix, and the transport is reduced by some factor $\delta$ (say). The previous formula is designed to interpolate smoothly between these two extremes as $\xi_1$ varies. 

Equations~(\ref{e33}) and (\ref{e34}) can be combined to give
\begin{equation}\label{e35}
\xi_0=\lambda \,\ln\left(\frac{1-\delta}{\lambda-\delta}\right).
\end{equation}
If follows that $\delta\leq \lambda \leq 1$, with $\xi_0=0$ when $\lambda=1$, and $\xi_0\rightarrow\infty$ as $\lambda \rightarrow \delta$. 
It is easily demonstrated that the function $\xi_0(\lambda)$ has a point of inflection when 
$\delta=\delta_{\rm crit} = 1/(1+{\rm e}^{\,2})= 0.1192$. This point corresponds to $\xi_0=4\,\delta_{\rm crit}= 0.4768$  and $\lambda=2\,\delta_{\rm crit}=0.2384$. 

\begin{figure}
\epsfysize=4in
\centerline{\epsffile{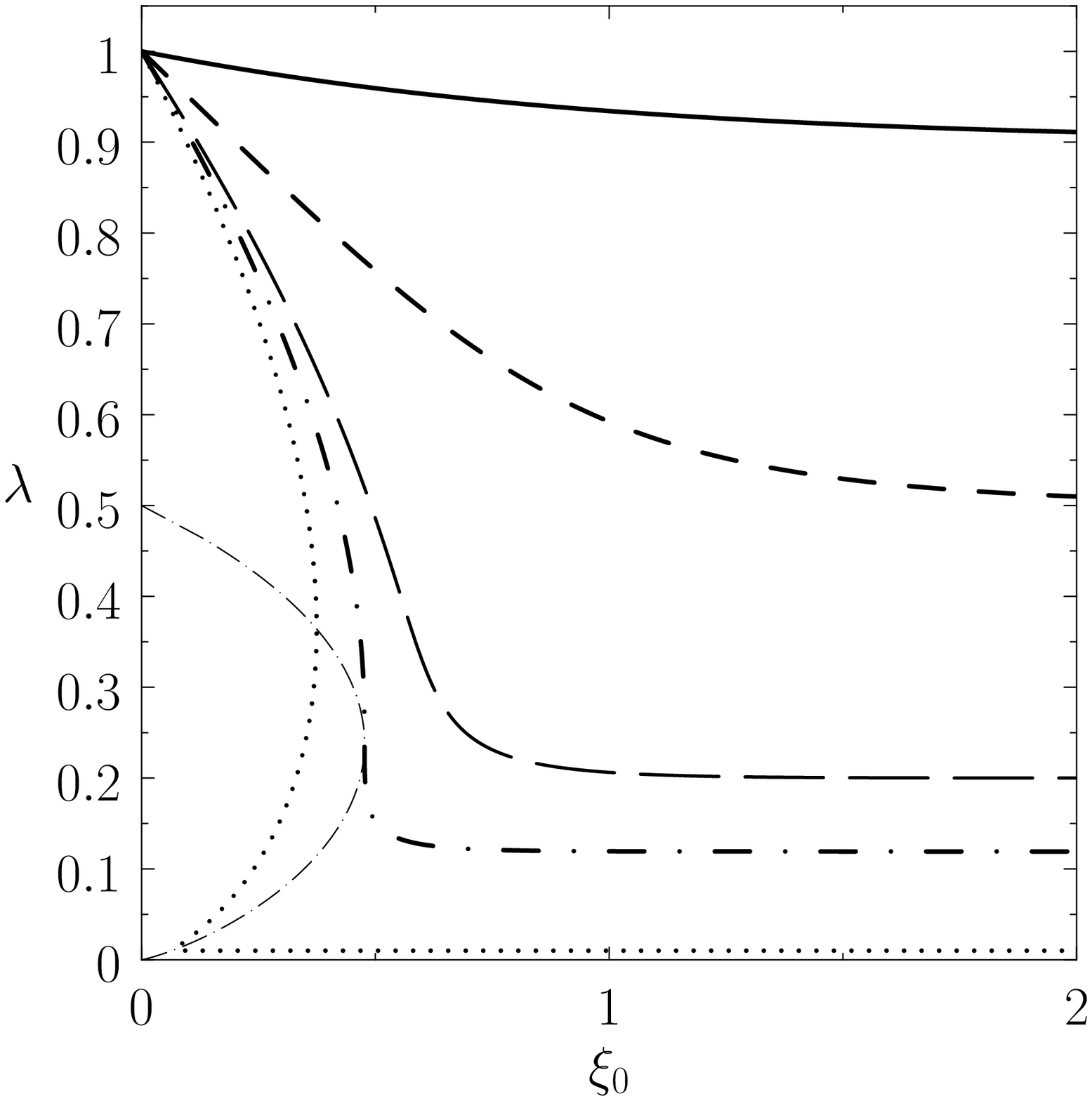}}
\caption{The perpendicular electron transport reduction parameter, $\lambda$,  plotted as a function of the  ratio of the divergences of the
parallel to perpendicular electron heat fluxes outside the island separatrix, $\xi_0$. The solid,  short-dashed, long-dashed, dot-short-dashed, and
dotted curves correspond to $\delta = 0.9$, $0.5$, $0.2$, $0.1192$, and $0.01$, respectively. Here, $\delta$ is the
maximum transport reduction parameter. 
The thin dot-long-dashed curve shows the locus of points where $d\xi_0/d\lambda=0$ ($d\xi_0/d\lambda>0$ to the
left of the curve, and vice versa).}\label{f1}
\end{figure}

Figure~\ref{f1} shows the perpendicular electron transport reduction parameter, $\lambda$, plotted as a function of the  ratio of the divergences of the
parallel to perpendicular electron heat fluxes outside the island separatrix, $\xi_0$, for various values of the
maximum transport reduction parameter, $\delta$. It can be seen that if $\delta>\delta_{\rm crit}$ then the $\xi_0$--$\lambda$ 
curves are such that $d\xi_0/d\lambda<0$ for  $\delta\leq \lambda \leq 1$. This implies that $\lambda$ decreases smoothly and continuously as $\xi_0$ increases, and vice versa. We shall refer to these solutions as {\em continuous solutions}\/ of Eq.~(\ref{e35}). 
On the other hand, if $\delta <\delta_{\rm crit}$ then the 
 $\xi_0$--$\lambda$ 
curves are such that $d\xi_0/d\lambda>0$ for  some intermediate range of $\lambda$ values lying between $\delta$ and $1$. 

\begin{figure}
\epsfysize=4in
\centerline{\epsffile{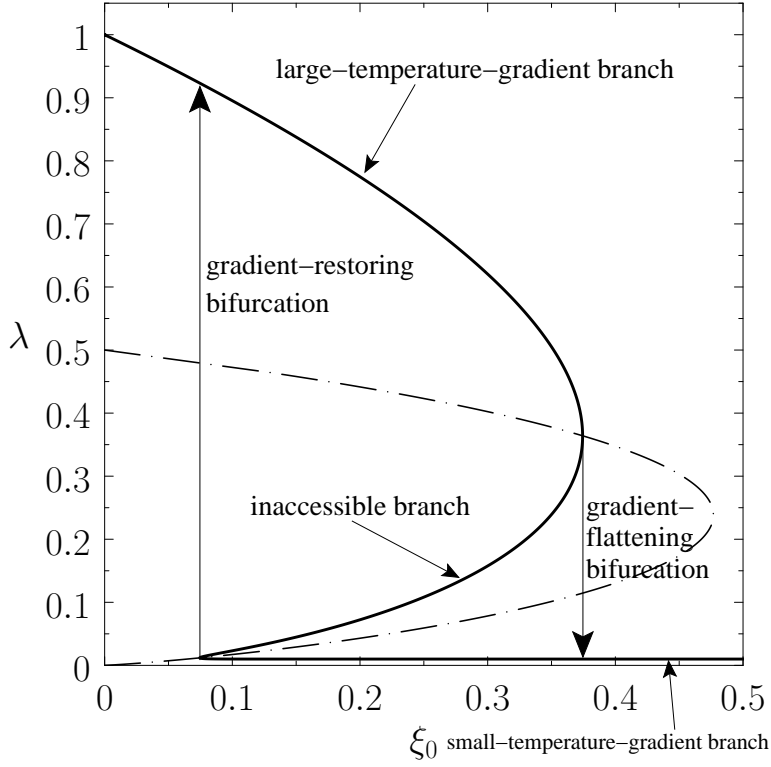}}
\caption{The solid curve shows the perpendicular electron transport reduction parameter, $\lambda$,  plotted as a function of the  ratio of the divergences of the
parallel to perpendicular electron heat fluxes outside the island separatrix, $\xi_0$, for $\delta =0.01$. Here, $\delta$ is the
maximum transport reduction parameter. 
The dot-long-dashed curve shows the locus of points where $d\xi_0/d\lambda=0$ ($d\xi_0/d\lambda>0$ to the
left of the curve, and vice versa). The various solution branches and bifurcations are labeled.}\label{f2}
\end{figure}

As illustrated in Fig.~\ref{f2}, the fact that  if  $\delta <\delta_{\rm crit}$ then $d\xi_0/d\lambda>0$  for intermediate values of $\lambda$ implies that there are two separate branches of solutions to Eq.~(\ref{e35})---the first characterized by $d\xi_0/d\lambda <0$ and relatively large $\lambda$, and the second characterized by $d\xi_0/d\lambda< 0$ and relatively small $\lambda$. We shall refer to the former solution branch as the {\em large-temperature-gradient branch}\/
[because it is characterized by a relatively large value of $\lambda$, which, from Eq.~(\ref{e34}), implies a relatively small value of $\xi_1$, which, from Eq.~(\ref{e41}), implies a relatively large electron temperature gradient inside the separatrix], and the latter as the {\em small-temperature-gradient branch}\/ [because it is characterized by a relatively small value of $\lambda$, which, from Eq.~(\ref{e34}), implies a relatively large value of $\xi_1$, which, from Eq.~(\ref{e41}), implies a relatively small electron temperature gradient inside the separatrix]. The two solution branches are separated by a dynamically inaccessible branch characterized
by $d\xi_0/d\lambda >0$. We shall refer to this branch of solutions as the {\em inaccessible branch}.  Referring to  Fig.~\ref{f2}, as $\xi_0$ increases from zero, we start off on the
large-temperature-gradient solution branch, and $\lambda$ decreases smoothly. However, when a critical value of
$\xi_0$ is reached (at which $d\xi_0/d\lambda=0$) there is a bifurcation to the small-temperature-gradient solution branch. 
We shall refer to this bifurcation as the {\em temperature-gradient-flattening bifurcation}, because it is characterized by
a sudden decrease in the transport ratio parameter, $\lambda$, which implies a sudden decrease in the electron
temperature gradient within the island separatrix. Once on  the small-temperature-gradient solution
branch, the control parameter $\xi_0$ must be reduced significantly in order to trigger a bifurcation back to the large-temperature-gradient solution  branch. 
We shall refer to this bifurcation as the {\em temperature-gradient-restoring bifuration}, because it is characterized by
a sudden increase in the transport ratio parameter, $\lambda$, which implies a sudden increase in the electron
temperature gradient within the island separatrix

\begin{figure}
\epsfysize=4in
\centerline{\epsffile{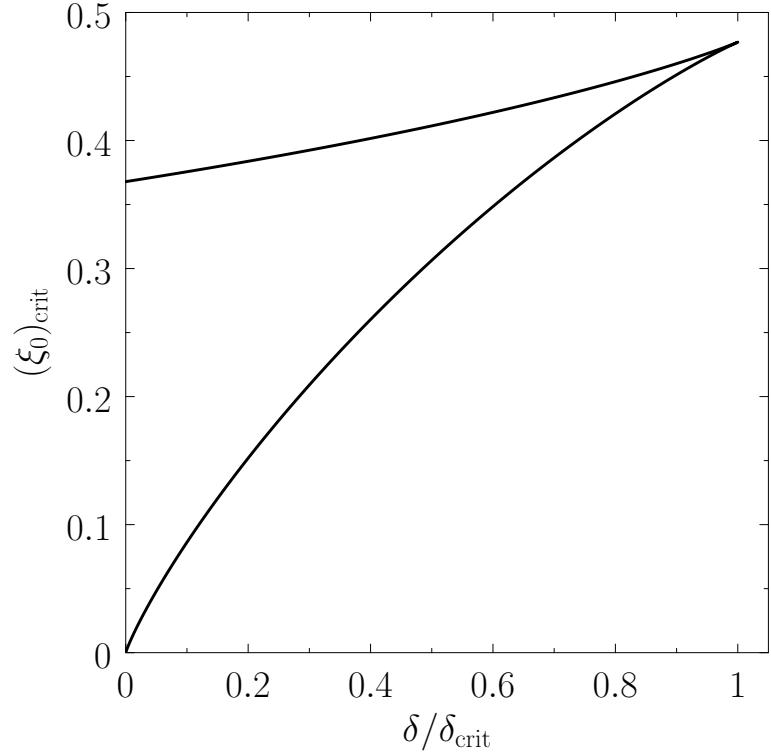}}
\caption{The  upper curve shows the critical value of the ratio of the divergences of the
parallel to perpendicular electron heat fluxes outside the island separatrix, $\xi_0$, below which a temperature-gradient-flattening bifurcation is triggered, plotted as a function of $\delta/\delta_{\rm crit}$.
The lower curve shows the critical value of  $\xi_0$ above which a temperature-gradient-restoring bifurcation is triggered, plotted as a function of $\delta/\delta_{\rm crit}$. Here, $\delta$ is the
maximum transport reduction parameter, and $\delta_{\rm crit}=0.1192$ is the critical value of $\delta$
below which bifurcations occur. 
}\label{f3}
\end{figure}

Figure~\ref{f3} shows the critical values
of the control parameter $\xi_0$ below and above which a temperature-gradient-flattening  and a temperature-gradient-restoring  bifurcation, respectively, are triggered, plotted
as a function of $\delta/\delta_{\rm crit}$.  

\begin{figure}
\epsfysize=4in
\centerline{\epsffile{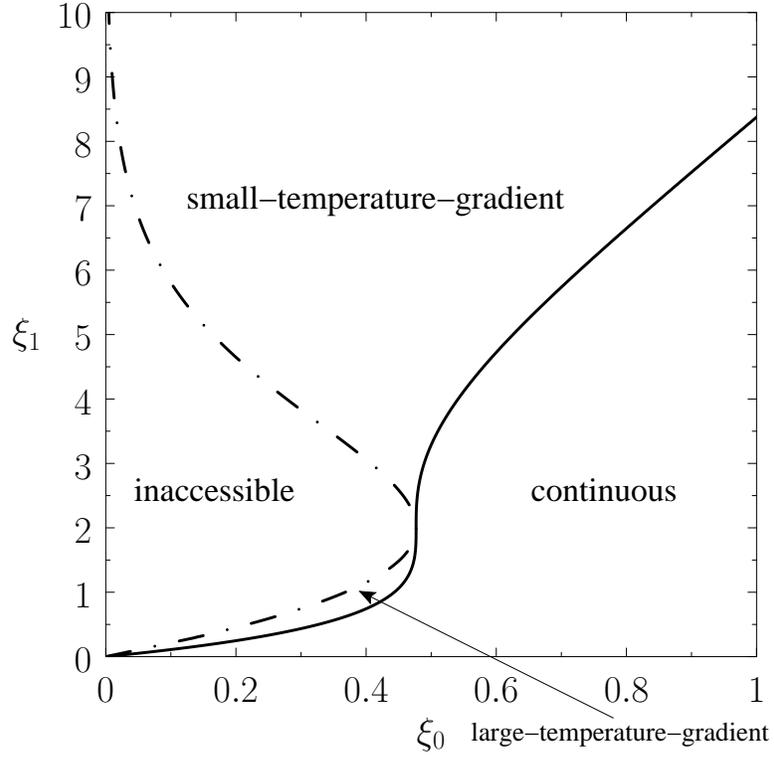}}
\caption{The  extents of the various solution branches of Eq.~(\ref{e35}) plotted in $\xi_0$--$\xi_1$ space. Here, $\xi_0$ is the ratio of the divergences of the
parallel to perpendicular electron heat fluxes outside the island separatrix,   whereas $\xi_1$  is the  same ratio inside the separatrix.}\label{f4}
\end{figure}

Figure~\ref{f4} shows the extents of the various solution branches (i.e., the continuous, large-temperature-gradient,
small-temperature-gradient, and inaccessible branches) plotted in $\xi_0$--$\xi_1$ space. 
It is clear that the large-temperature-gradient solution branch
is characterized by $\xi_0\ll 1$ and $\xi_1\ltapp 1$. In other words, the region outside the island
separatrix lies in the narrow-island limit, $W\ll W_{c\,0}$, whereas that inside the separatrix lies in the
narrow/intermediate island limit, $W\ltapp W_{c\,1}$. [See Eqs.~(\ref{e31}) and (\ref{e32}).]  This implies weak to moderate flattening of the
temperature gradient  within the separatrix. 
On the other hand, the small-temperarture-gradient
solution branch is characterized by $\xi_0\ll 1$ and $\xi_1\gg 1$. In other words, the region
outside the island separatrix lies in the narrow-island limit,  $W\ll W_{c\,0}$, whereas that
inside the separatrix lies in the wide-island limit, $W\gg W_{c\,1}$.  This implies  strong flattening of
the temperature gradient  within the separatrix. Figure~\ref{f4} suggests that bifurcated solutions of Eq.~(\ref{e35}) occur because it is
possible for the regions inside and outside the island separatrix to lie in opposite asymptotic limits
(the two possible limits being the wide-island and the narrow-island limits). Obviously, this is not possible in the
conventional model in which $\kappa_\perp$ is taken to be spatially uniform in the island region. 

Finally, according to our simple model, the critical value of the maximum transport reduction
parameter, $\delta$, below which bifurcated solutions of the electron energy transport equation  occur is $0.1192$. As we have seen, there is
experimental evidence for a transport reduction within the separatrix of a neoclassical island chain
by between 1 and 2 orders of magnitude.\cite{ex5} According to our model, such a reduction would be  large enough to  generate bifurcated solutions. 

\subsection{Composite Island Temperature Profile Model}
Let
\begin{equation}
\delta T_{\rm narrow}(X,\zeta,\xi) = X + \xi^{1/4}\,\frac{\sqrt{2}}{4}\,f\left(\sqrt{2}\,\xi^{1/4}\, X\right)\,\cos\zeta
\end{equation}
be the island temperature profile in the narrow-island limit. [See Eq.~(\ref{e14}).] Here, $\xi=(W/W_c)^{\,4}$. 
Likewise, let 
\begin{equation}
\delta T_{\rm wide}(X,\zeta) = \left\{
\begin{array}{ccc}
0&\mbox{\hspace{1cm}}& -1\leq \psi\leq 1\\[0.5ex]
s\int_0^\psi \frac{d\psi'}{\langle X^{\,2}\rangle (\psi')}&&\psi>1\end{array}\right..
\end{equation}
be the island temperature profile in the wide-island limit.  [See Eqs.~(\ref{e23}) and (\ref{e24}).]
Let us write
\begin{equation}
\delta T(X,\zeta) = \left\{
\begin{array}{ccc}
\delta T_{\rm inside}(X,\zeta)&\mbox{\hspace{1cm}}&-1\leq \psi\leq 1\\[0.5ex]
\delta T_{\rm outside}(X,\zeta)&&\psi>1\end{array}\right.,
\end{equation}
where [cf., Eq.~(\ref{e34})] 
\begin{equation}
\delta T_{\rm inside}(X,\zeta)= {\rm e}^{-\xi_1}\,\delta T_{\rm narrow}(X,\zeta,\xi_1)+(1-{\rm e}^{-\xi_1})\,\delta T_{\rm wide}(X,\zeta)
\end{equation}
and 
\begin{equation}
\delta T_{\rm outside}(X,\zeta)= {\rm e}^{-\xi_0}\,\delta T_{\rm narrow}(X,\zeta,\xi_0)+(1-{\rm e}^{-\xi_0})\,\delta T_{\rm wide}(X,\zeta).
\end{equation}
It follows that
\begin{align}\label{e41}
\frac{\partial \,\delta T_{\rm inside}}{\partial X}&\simeq {\rm e}^{-\xi_1}\left[1+ \frac{f'(0)}{2}\,\xi_1^{\,1/2}\,\cos\zeta+ {\cal O}(\xi_1)\right],\\[0.5ex]
\frac{\partial\,\delta T_{\rm outside}}{\partial X}&\simeq {\rm e}^{-\xi_0}\left[1+ \frac{f'(0)}{2}\,\xi_0^{\,1/2}\,\cos\zeta+{\cal O}(\xi_0)\right]
+(1-{\rm e}^{-\xi_0})\,\frac{X}{\langle X^{\,2}\rangle}.\label{e42}
\end{align}
Here, $f'(0)=1.1981$, as determined from the numerical solution of Eq.~(\ref{e13}). 

\subsection{Evaluation of Integrals}
According to Eqs.~(\ref{e26a}), (\ref{e27}), (\ref{e41}), and (\ref{e42}), 
\begin{align}\label{e72}
G_1 &= 2\,(I_2+ I_3),\\[0.5ex]
G_2 &= 16\,I_1\,({\rm e}^{-\xi_0}-{\rm e}^{-\xi_1})+ 8\,f'(0)\,I_2\,\xi_1^{\,1/2}\,{\rm e}^{-\xi_1}
+ 8\,f'(0)\,I_3\,\xi_0^{\,1/2}\,{\rm e}^{-\xi_0}\nonumber\\[0.5ex]
&\phantom{=}+16\,I_4\,(1-{\rm e}^{-\xi_0}),\label{e73}
\end{align}
where
\begin{align}\label{e45}
I_1 &= -\int_{-1}^1\langle \cos\zeta\rangle\,d\psi,\\[0.5ex]
I_2 &= \int_{-1}^{1}\frac{\langle \cos\zeta\rangle^{\,2}}{\langle 1\rangle}\,d\psi,\\[0.5ex]
I_3&= \int_{1}^\infty\frac{\langle \cos\zeta\rangle^{\,2}}{\langle 1\rangle}\,d\psi,\\[0.5ex]
I_4&=\int_1^{\infty} \frac{\langle \cos\zeta\rangle}{\langle X^{\,2}\rangle\,\langle 1\rangle}\,d\psi.\label{e48}
\end{align}
Here,  use has been made of the easily proved result 
\begin{equation}
 \int_{-1}^\infty \langle \cos\zeta\rangle\,d\psi = 0.
 \end{equation}
 
 Let $\psi=2\,k^{\,2}-1$. It follows that $d\psi =4\,k\,dk$. In the region $0\leq k \leq 1$, we
can write
\begin{align}\label{e50}
\zeta &= 2\,\cos^{-1}(k\,\sin\vartheta),\\[0.5ex]
X &= 2\,k\,\cos\vartheta,\\[0.5ex]
\cos\zeta&= 1-2\,(1-k^{\,2}\,\sin^2\vartheta),\\[0.5ex]
\langle A\rangle &=\int_{-\pi/2}^{\pi/2}\frac{A(k,\vartheta)}{\sqrt{1-k^{\,2}\,\sin^2\vartheta}}\,\frac{d\vartheta}{2\pi}.
\end{align}
On the other hand, in the region $k>1$, we can write
\begin{align}
\zeta&=\pi-2\,\vartheta,\\[0.5ex]
X &= 2\sqrt{k^{\,2}-\sin^2\vartheta},\\[0.5ex]
\cos\zeta &= 2\,k^{\,2}-1-2\,(k^{\,2}-\sin^2\vartheta),\\[0.5ex]
\langle A\rangle &=\int_{-\pi/2}^{\pi/2}\frac{A(k,\vartheta)}{\sqrt{k^{\,2}-\sin^2\vartheta}}\,\frac{d\vartheta}{2\pi}.\label{e57}
\end{align}
Here, it is assumed that $A$ is an even function of $X$. 

Let
\begin{align}\label{e58}
{\cal A}(k)&= 2\,k\,\langle 1\rangle,\\[0.5ex]
{\cal B}(k)&=2\,k\,\langle\cos\zeta\rangle,\\[0.5ex]
{\cal C}(k) &=\frac{\langle X^{\,2}\rangle}{2\,k}.\label{e60}
\end{align}
It follows from Eqs.~(\ref{e50})--(\ref{e57}) that in the region $0\leq k \leq 1$,
\begin{align}
{\cal A}(k) &= \frac{2}{\pi}\,k\,K(k),\\[0.5ex]
{\cal B}(k) &=\frac{2}{\pi}\,k\left[K(k)-2\,E(k)\right].
\end{align}
On the other hand, in the region $k>1$, 
 \begin{align}
{\cal A}(k) &= \frac{2}{\pi}\,K(1/k),\\[0.5ex]
{\cal B}(k) &=\frac{2}{\pi}\left[(2\,k^{\,2}-1)\,K(1/k) -2\,k^{\,2}\,E(1/k)\right],\\[0.5ex]
{\cal C}(k) &= \frac{2}{\pi}\,E(1/k).
\end{align}
Here,
\begin{align}
K(k) &=\int_0^{\pi/2} (1-k^{\,2}\,\sin^2\vartheta)^{-1/2}\,d\vartheta,\\[0.5ex]
E(k) &= \int_0^{\pi/2} (1-k^{\,2}\,\sin^2\vartheta)^{1/2}\,d\vartheta
\end{align}
are complete elliptic integrals.\cite{elliptic}
Hence, according to Eqs.~(\ref{e45})--(\ref{e48}) and (\ref{e58})--(\ref{e60}), 
\begin{align}
I_1 &=-2 \int_0^1 {\cal B}\,dk=0.4244,\\[0.5ex]
I_2 &= 2\int_0^1\frac{{\cal B}^{\,2}}{\cal A}\,dk=0.3527,\\[0.5ex]
I_3&= 2\int_1^\infty\frac{{\cal B}^{\,2}}{\cal A}\,dk=0.0587,\\[0.5ex]
I_4 &= 2\int_1^\infty \frac{{\cal B}}{{\cal A}\,{\cal C}}\,dk=0.3838.
\end{align}
Thus, Eqs.~(\ref{e72}) and (\ref{e73}) yield 
\begin{align}
G_1 &=0.8227, \\[0.5ex]
G_2 &= 6.791\,({\rm e}^{-\xi_0}-{\rm e}^{-\xi_1})+3.380\,\xi_1^{\,1/2}\,{\rm e}^{-\xi_1}+0.562\,\xi_0^{\,1/2}\,{\rm e}^{-\xi_0}+6.140\,(1-{\rm e}^{-\xi_0}),\label{e73a}
\end{align}
respectively. 

\subsection{Destabilizing Effect of Perturbed Bootstrap Current}\label{se}
\begin{figure}
\epsfysize=4in
\centerline{\epsffile{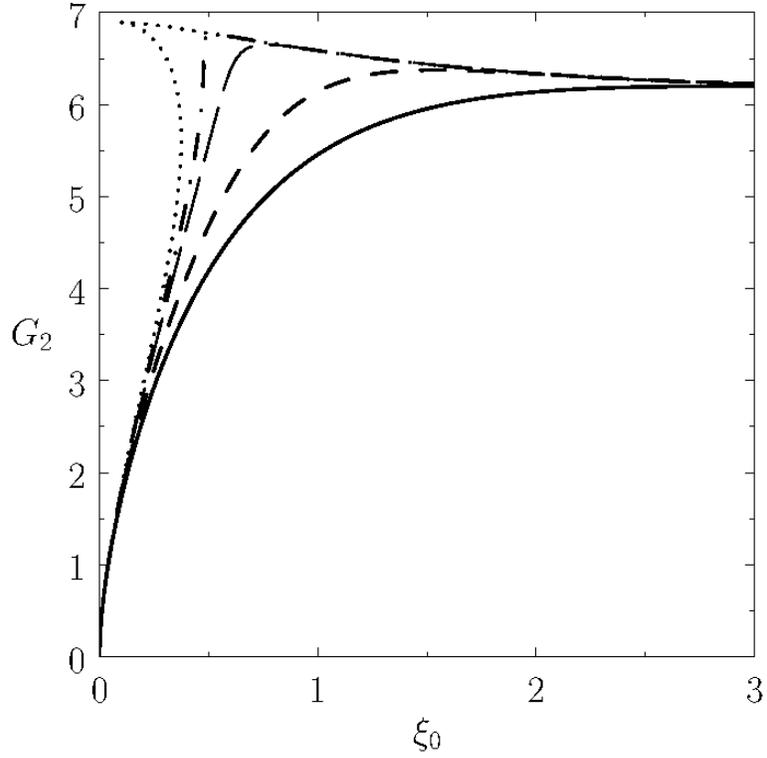}}
\caption{The bootstrap destabilization parameter, $G_2$, plotted as a function of the  ratio of the divergences of the
parallel to perpendicular electron heat fluxes outside the island separatrix, $\xi_0$. The solid,  short-dashed, long-dashed, dot-short-dashed, and
dotted curves correspond to $\delta = 0.9$, $0.5$, $0.2$, $0.1192$, and $0.01$, respectively. Here, $\delta$ is the
maximum transport reduction parameter. 
}\label{f5}
\end{figure}

The dimensionless parameter $G_2$, appearing in the modified Rutherford equation, (\ref{e26}), measures the
destabilizing influence of the perturbed bootstrap current. 
Figure~\ref{f5} shows $G_2$ plotted as a function of the so-called {\em neoclassical tearing mode control parameter}, 
\begin{equation}\label{e74}
\xi_0 = \left(\frac{W}{W_{c\,0}}\right)^4 = \left(\frac{W}{4}\right)^4\left(\frac{\kappa_\parallel}{\kappa_{\perp\,0}}\right)
\left(\frac{k_\theta}{L_s}\right)^{2},
\end{equation}
which measures the ratio of the divergences of the parallel to the perpendicular electron heat fluxes outside the island
separatrix.  [See Eqs.~(\ref{e29}) and (\ref{e31}).] The curves shown in this figure are obtained from  Eqs.~(\ref{e33}), (\ref{e35}), and (\ref{e73a}). Note that $\kappa_\parallel$ and $\kappa_{\perp\,0}$  are the local (to the rational surface) parallel and perpendicular
electron thermal conductivities, respectively, in the absence of an island chain. 

It can be seen, from Fig.~\ref{f5}, that if the maximum transport reduction parameter, 
$\delta$, is relatively close to unity (implying a relatively weak reduction in the perpendicular electron thermal conductivity
inside the island separatrix when the electron temperature profile is completely flattened in this region) then the bootstrap
destabilization parameter, $G_2$, increases monotonically with increasing $\xi_0$, taking the value
$3.492\,\xi_0^{\,1/2}$ when $\xi_0\ll 1$, and approaching the value $6.140$ asymptotically as $\xi_0\rightarrow\infty$.\cite{rf}  [These two  limits follow from Eq.~(\ref{e73a}), given that
$\xi_1\simeq \xi_0$ when $\lambda\simeq 1$.]

According to Fig.~\ref{f5}, as $\delta$ decreases significantly below unity (implying an increasingly strong reduction in the perpendicular electron thermal conductivity
inside the island separatrix when the electron temperature profile is completely flattened in this region) it remains the
case that $G_2=3.492\,\xi_0^{\,1/2}$ when $\xi_0\ll 1$, and $G_2\rightarrow 6.140$ as $\xi_0\rightarrow\infty$. 
However, at intermediate values of $\xi_0$ [i.e., $\xi_0 \sim{\cal O}(1)$], the rate of increase of $G_2$ with
$\xi_0$ becomes increasingly steep. This result suggests that a substantial reduction in the perpendicular electron
thermal conductivity inside the island separatrix, when the electron temperature profile is completely flattened in this
region, causes the bootstrap destabilization term in the modified Rutherford equation, (\ref{e26}), to ``switch
on''  much more rapidly as the neoclassical tearing mode control parameter, $\xi_0$, is increased, compared to the standard case in which there is no reduction in the conductivity.

\begin{figure}
\epsfysize=4in
\centerline{\epsffile{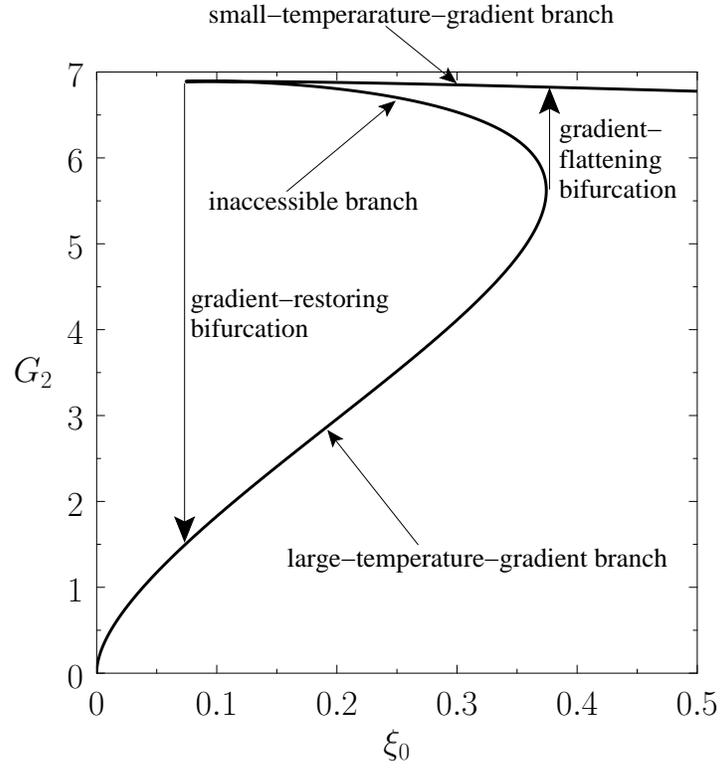}}
\caption{The bootstrap destabilization parameter, $G_2$,  plotted as a function of the  ratio of the divergences of the
parallel to perpendicular electron heat fluxes outside the island separatrix, $\xi_0$, for $\delta =0.01$. Here, $\delta$ is the
maximum transport reduction parameter. 
The various solution branches and bifurcations are labeled.}\label{f6}
\end{figure}

Finally, it is apparent from Fig.~\ref{f5} that if $\delta$ falls below the critical value $\delta_{\rm crit}=0.1192$ then
the bootstrap destabilization parameter, $G_2$, becomes a multi-valued function of $\xi_0$ at intermediate values
of $\xi_0$. As illustrated in Fig.~\ref{f6}, this behavior is due to the existence of separate branches of solutions
of the electron energy conservation equation. (See Sect.~\ref{sxx}.) The large-temperature-gradient branch
is characterized by relatively weak flattening of the electron temperature profile within the island
separatrix, and a consequent  relatively small value (i.e., significantly smaller than the asymptotic limit $6.140$) of the bootstrap destabilization parameter, $G_2$. On the other hand, the small-temperature-gradient branch is
characterized by almost complete flattening of the electron  temperature profile within the island
separatrix. Consequently, the bootstrap destabilization parameter, $G_2$, takes a value close to the
asymptotic limit $6.140$ on this solution branch. The large-temperature-gradient and small-temperature-gradient
solution branches are separated by a dynamically inaccessible branch of solutions.  Referring to Fig.~\ref{f6}, as the neoclassical tearing mode control parameter, $\xi_0$, increases from a value much less than unity, we start off on the large-temperature-gradient
solution branch, and the bootstrap destabilization parameter, $G_2$, increases smoothly and monotonically
from a small value. However, when a critical value of $\xi_0$ is reached, there is a gradient-flattening-bifurcation
to the small-temperature-gradient solution branch. This bifurcation is accompanied by a sudden increase in $G_2$ to a value close to its asymptotic limit $6.140$. 
Once on the small-temperature-gradient solution branch, $\xi_0$ must be decreased by a significant amount
before a gradient-restoring-bifurcation to the large-temperature-gradient solution branch is triggered. 
Moreover, the gradient-restoring-bifurcation is accompanied by a very large reduction in $G_2$. 

\subsection{Long Mean-Free-Path Effects}
The parallel electron thermal conductivity takes the form\,\cite{brag}
\begin{equation}\label{e75}
\kappa_\parallel \sim n_e\,v_e\,\lambda_e
\end{equation}
in a collisional plasma,  where $n_e$ is the electron number density, $v_e$ is the electron themal
velocity, and $\lambda_e$ is the electron mean-free-path.  However, in a conventional tokamak
plasma the mean-free-path $\lambda_e$ typically exceeds the parallel (to the magnetic field) wavelength
$\lambda_\parallel$ of low-mode-number helical perturbations. Under these circumstances, the simple-minded application of Eq.~(\ref{e75}) yields
unphysically large parallel heat fluxes. The parallel conductivity in the physically-relevant long-mean-free-path
limit ($\lambda_e\ll \lambda_\parallel$) can be crudely estimated as\,\cite{rf,gor}
\begin{equation}
\kappa_\parallel \sim n_e\,v_e\,\lambda_\parallel,
\end{equation}
which is equivalent to replacing parallel conduction by parallel convection in the electron energy conservation
equation, (\ref{e1}). For a magnetic island of full radial width $W$, the typical value of $\lambda_\parallel$ is
$n_\varphi\,s_s\,w/R_0$. Hence, in the long-mean-free-path limit, the expression for the neoclassical
tearing mode control parameter (\ref{e74}) is replaced by
\begin{equation}\label{e77}
\xi_0  = \left(\frac{W}{4}\right)^5\left(\frac{\kappa_\parallel'}{\kappa_{\perp\,0}}\right)
\left(\frac{k_\theta}{L_s}\right)^{2},
\end{equation}
where $\kappa_\parallel' = n_\varphi\,n_e\,v_e\,s_s/R_0$, and $n_e$ and $v_e$ are evaluated at the rational surface. 

\section{Summary and Discussion}
In this paper, we have investigated the effect of the reduction in anomalous perpendicular electron heat transport inside
the separatrix of a magnetic island chain associated with a neoclassical tearing mode in a tokamak plasma, due to the flattening of the electron temperature
profile in this region, on the overall stability of the mode. Our model (which is described in Sect.~\ref{s3}) is fairly crude, in that the
perpendicular electron thermal conductivity, $\kappa_\perp$,  is simply assumed to take different spatially-uniform values
in the regions inside and outside the separatrix. Moreover, when the temperature profile is completely flattened within the
island separatrix, $\kappa_\perp$ in this region is assumed to be reduced by some factor $\delta$, where $0<\delta\leq 1$. The degree of temperature flattening inside the separatrix is ultimately controlled by a dimensionless
parameter $\xi_0$ that measures the ratio of the divergences of the parallel and perpendicular electron heat
fluxes in the vicinity of the island chain. Expressions for $\xi_0$ in the short-mean-free-path and the more 
physically-relevant long-mean-free-path limits are given in Eqs.~(\ref{e74})  and (\ref{e77}), respectively. Finally, the destabilizing influence of the perturbed bootstrap current is parameterized in terms of a dimensionless quantity $G_2>0$ that appears in the modified
Rutherford equation. [See Eqs.~(\ref{e26}) and (\ref{e27}).] A large value of $G_2$ implies
substantial destabilization, and vice versa. 

In the standard case $\delta=1$ (in which there is no reduction in the perpendicular
electron thermal conductivity inside the island separatrix when the electron temperature
profile is completely flattened in this region), the bootstrap destabilization parameter $G_2$
increases smoothly and monotonically as the control parameter $\xi_0$ increases, from a
value much less than unity when $\xi_0\ll 1$, to the  asymptotic limit $6.140$ when $\xi_0\gg 1$.\cite{rf} (See Section~\ref{se}.)

As $\delta$ decreases significantly below unity (implying an increasingly
strong reduction in the perpendicular
electron thermal conductivity inside the island separatrix when the electron temperature
profile is completely flattened in this region), the small-$\xi_0$ and large-$\xi_0$ behaviors
of the bootstrap destabilization parameter remain unchanged. However, at intermediate
values of the control parameter $\xi_0$  [i.e., $\xi_0\sim {\cal O}(1)$], the rate of increase of
$G_2$ with $\xi_0$ becomes increasingly steep. (See Fig.~\ref{f5}.) In other words, a substantial
reduction in the perpendicular electron thermal conductivity inside the island separatrix, when the electron
temperature profile is completely flattened in this region, causes the bootstrap destabilization
parameter $G_2$ to ``switch on" much more rapidly as the control parameter $\xi_0$ is increased,
compared to the standard case in which $\delta=1$. (See Section~\ref{se}.)

Finally, if $\delta$ falls below the critical value  $0.1192$ then the bootstrap
destabilization parameter, $G_2$, becomes a multi-valued function of the control
parameter $\xi_0$, at intermediate values of $\xi_0$. This behavior is due to the existence of separate branches of solutions
of the electron energy conservation equation. (See Sect.~\ref{sxx}.) The {\em large-temperature-gradient branch}\/
is characterized by relatively weak flattening of the electron temperature profile within the island
separatrix, and a consequent  relatively small value (i.e., significantly smaller than the asymptotic limit $6.140$) of the bootstrap destabilization parameter, $G_2$. On the other hand, the {\em small-temperature-gradient branch}\/ is
characterized by almost complete flattening of the electron  temperature profile within the island
separatrix. Consequently, the bootstrap destabilization parameter, $G_2$, takes a value close to the
asymptotic limit $6.140$ on this solution branch.  The large-temperature-gradient and small-temperature-gradient
solution branches are separated by a dynamically inaccessible branch of solutions. As the  control parameter, $\xi_0$, increases from a value much less than unity, the system starts off on the large-temperature-gradient
solution branch, and the bootstrap destabilization parameter, $G_2$, increases smoothly and monotonically
from a small value. However, when a critical value of $\xi_0$ is reached, there is a {\em gradient-flattening-bifurcation}\/
to the small-temperature-gradient solution branch. (See Fig.~\ref{f6}.) This bifurcation is accompanied by a sudden increase in $G_2$ to a value close to its asymptotic limit $6.140$. 
Once on the small-temperature-gradient solution branch, $\xi_0$ must be decreased by a significant amount
before a {\em gradient-restoring-bifurcation}\/ to the large-temperature-gradient solution branch is triggered. 
Moreover, the gradient-restoring-bifurcation is accompanied by a very large reduction in $G_2$. (See Section~\ref{se}.) 

The behavior  described  in the preceding paragraph points to the disturbing possibility that a neoclassical tearing mode in a tokamak plasma could become essentially {\em self-sustaining}. In other words, once the mode is triggered, and the electron temperature profile
is flattened within the island separatrix, the consequent substantial reduction in the perpendicular thermal conductivity in this region reinforces the temperature flattening, making
it very difficult to remove the mode from the plasma. 
 
\section*{Acknowledgements}
This research was funded by the U.S.\ Department of Energy under contract DE-FG02-04ER-54742.

\newpage

\newpage

\end{document}